\documentstyle[12pt]{article}
\begin{document}
\hfuzz=1pt
\setlength{\textheight}{8.5in}
\setlength{\topmargin}{0in}
\begin{centering}
\LARGE {\bf Quantum oblivious transfer protocols based on EPR
states} \\  \vspace{.75in}
\Large M. Ardehali
\footnote[1] {Present and permanent address:
Microelectronics Research Laboratories,
NEC Corporation,
1120 Shimokuzawa,
Sagamihara,
Kanagawa 229
Japan. Correspondence concerning this manuscript should be
sent to this address.}
\\ \vspace{.5in}
\large 4-31-12, ATAGO Center, Tama-shi, Tokyo 206 Japan
\\ \vspace{.75in}
\end{centering}

We describe efficient protocols for quantum oblivious transfer
and for one-out-of-two quantum oblivious transfer.
These protocols, which can be implemented with present
technology, are secure against general attacks as long as the cheater
can not store the bit for
an arbitrarily long period of time.

\pagebreak

In 1970, Wiesner \cite{1}
wrote a highly innovative paper about quantum
cryptography \cite{2}, \cite{3}.
In his paper, he also introduced the concept of
{\em Multiplexing}, which was later rediscovered by Rabin, \cite{4}
and is now
usually called {\em Oblivious Transfer} (OT).
Let us briefly describe the OT protocol:
\\
1 - Alice knows one bit $\lambda$, where $\lambda$ is either $1$ or
$-1$ \cite{5}.
\\
2 - Bob obtains bit $\lambda$ from Alice with probability 0.5.
\\
3 - Bob knows whether or not he obtained bit $\lambda$.
\\
4 - Alice does not learn whether or not Bob obtained bit $\lambda$.

In this letter, we propose a very efficient  protocol for
quantum oblivious transfer which is
secure even against cheaters with
unlimited computing power, if we assume that Bob can not
store the bit for an arbitrarily long period of time.
It is worth noting that none of the known
non-quantum protocols for oblivious transfer
are perfectly secure; they all allow one among Alice or Bob to cheat
without risk of detection if he
or she can break some unproved cryptographic assumptions.

Before proceeding, it is useful
to review some elementary features of
quantum mechanics. We consider an
unstable source emitting pairs of
entangled Einstein, Podolsky, Rosen (EPR) \cite{8} particles.
We take the $z$ axis along the direction
of the flight of the particles, and the
$x$ and the $y$ axes along any two
directions perpendicular to the $z$ axis.
For a pair of particles in the EPR state
$\mid\!\phi\rangle=\frac{\displaystyle 1}
{\displaystyle \sqrt{2}}\left(\mid\uparrow
\uparrow\rangle+\mid\downarrow\downarrow\rangle\right)$,
the expected value of the product
of the spin of the particles along
two arbitrary axes $\vec{a}$
and $\vec{b}$ in the $xy$ plane is \cite{9}

\begin{equation}
\langle\phi\mid\sigma_{1}^{a}\sigma_{2}^{b}\mid\phi\rangle=
\cos\left(\theta_{a}+\theta_{b}\right),
\end {equation}
where $\sigma_{1}^{a}$ is the spin of the first particle along
axis $\vec{a}$,
$\sigma_{2}^{b}$ is the spin of the second particle along
axis $\vec{b}$,
and $\theta_{a}\left(\theta_{b}\right)$ is the angle
between axis $\vec{a}\left(\vec{b}\right)$ and the $x$ axis.
Thus the spin of the first particle along the $x$ axis,
$m_{1}^{x}$,
is perfectly correlated with the spin of the second particle
along the same axis, i.e.,
$m_{1}^{x}m_{2}^{x}=1$.
However, the spin of the first particle along
the $y$ axis is perfectly anticorrelated with the spin of the second
particle along the same axis, i.e,
$m_{1}^{y}m_{2}^{y}=-1$.
Similarly for EPR states \cite{10}

\begin{eqnarray}
\mid\!\!\phi^\prime\rangle=\frac{\displaystyle 1}
{\displaystyle \sqrt {2}}\left(\mid\uparrow
\uparrow\rangle-\mid\downarrow\downarrow\rangle\right),& \qquad \quad
\langle\phi^\prime\mid\sigma_{1}^{a}
\sigma_{2}^{b}\mid\phi^\prime\rangle=
-\cos\left(\theta_{a}+\theta_{b}\right),\nonumber \\
\mid\!\psi\rangle=\frac{\displaystyle 1}
{\displaystyle \sqrt{2}}\left(\mid\uparrow
\downarrow\rangle+\mid\downarrow\uparrow\rangle\right),& \qquad \quad
\langle\psi\mid\sigma_{1}^{a}\sigma_{2}^{b}\mid\psi\rangle=
\cos\left(\theta_{a}-\theta_{b}\right),\nonumber \\
\mid\!\psi^\prime\rangle=\frac{\displaystyle 1}
{\displaystyle \sqrt{2}}\left(\mid\uparrow
\downarrow\rangle-\mid\downarrow\uparrow\rangle\right),& \qquad \quad
\langle\psi^\prime\mid\sigma_{1}^{a}
\sigma_{2}^{b}\mid\psi^\prime\rangle=
-\cos\left(\theta_{a}-\theta_{b}\right),\nonumber \\
\mid\!\alpha\rangle=\frac{\displaystyle 1}
{\displaystyle \sqrt{2}}\left(\mid\uparrow
\uparrow\rangle+i\mid\downarrow\downarrow\rangle\right),& \qquad \quad
\langle\alpha\mid\sigma_{1}^{a}\sigma_{2}^{b}\mid\alpha\rangle=
\sin\left(\theta_{a}+\theta_{b}\right),\\
\mid\!\alpha^\prime\rangle=\frac{\displaystyle 1}
{\displaystyle \sqrt{2}}\left(\mid\uparrow
\uparrow\rangle-i\mid\downarrow\downarrow\rangle\right),& \qquad \quad
\langle\alpha^\prime\mid\sigma_{1}^{a}
\sigma_{2}^{b}\mid\alpha^\prime\rangle=
-\sin\left(\theta_{a}+\theta_{b}\right),\nonumber \\
\mid\!\beta\rangle=\frac{\displaystyle 1}
{\displaystyle \sqrt{2}}\left(\mid\uparrow
\downarrow\rangle+i\mid\downarrow\uparrow\rangle\right),& \qquad \quad
\langle\beta\mid\sigma_{1}^{a}\sigma_{2}^{b}\mid\beta\rangle=
\sin\left(\theta_{a}-\theta_{b}\right),\nonumber \\
\mid\!\beta^\prime\rangle=\frac{\displaystyle 1}
{\displaystyle \sqrt{2}}\left(\mid\uparrow
\downarrow\rangle-i\mid\downarrow\uparrow\rangle\right),& \qquad \quad
\langle\beta^\prime\mid\sigma_{1}^{a}
\sigma_{2}^{b}\mid\beta^\prime\rangle=
-\sin\left(\theta_{a}-\theta_{b}\right).\nonumber \\
\nonumber
\end{eqnarray}

With the above in mind, we now proceed to describe
the following protocol:
\\
(1) Alice and Bob agree that the bit $\lambda$ is encoded in the
product of the spin of the first and the
second particles along the $x$ axis
\begin{math}
(m_{1}^{x}m_{2}^{x}
\end{math}
or
\begin{math}
-m_{1}^{x}m_{2}^{x}),
\end{math}
or the spin of the
first and the second particles
along the $y$ axis
\begin{math}
(m_{1}^{y}m_{2}^{y}
\end{math}
or
\begin{math}
-m_{1}^{y}m_{2}^{y}),
\end{math}
or the spin of the first particle along the
$y$ axis and the spin of the second particle along
the $x$ axis
\begin{math}
(m_{1}^{y}m_{2}^{x}
\end{math}
or
\begin{math}
-m_{1}^{y}m_{2}^{x}),
\end{math}
or the spin of the first particle along the $-x$ axis
and the spin of the second particle along the
$y$ axis (
\begin{math}
m_{1}^{-x}m_{2}^{y}
\end{math}
or
\begin{math}
-m_{1}^{-x}m_{2}^{y}).
\end{math}
They also agree
on a set $A$ defined as
$A=\{m_{1}^{x}m_{2}^{x},m_{1}^{y}m_{2}^{y}\}$, and on a set $B$
defined as
$B=\{{m_{1}^{y}m_{2}^{x},m_{1}^{-x}m_{2}^{y}}\}$.
\\
(2) Alice encodes the bit to be obliviously transferred in
\begin{math}
m_{1}^{x}m_{2}^{x},
\end{math}
or
\begin{math}
-m_{1}^{x}m_{2}^{x},
\end{math}
or
\begin{math}
m_{1}^{y}m_{2}^{y},
\end{math}
or
\begin{math}
-m_{1}^{y}m_{2}^{y},
\end{math}
or
\begin{math}
m_{1}^{y}m_{2}^{x},
\end{math}
or
\begin{math}
-m_{1}^{y}m_{2}^{x},
\end{math}
or
\begin{math}
m_{1}^{-x}m_{2}^{y},
\end{math}
or
\begin{math}
-m_{1}^{-x}m_{2}^{y},
\end{math}
chosen randomly by her.
She randomly chooses an appropriate state
(as shown below, there are two appropriate states
for any of her choices).
She then prepares a pair of particles in that state and sends
both particles to Bob.
\\
(3) Bob measures randomly either $m_{1}^{x}m_{2}^{x}$ or
$m_{1}^{y}m_{2}^{y}$, or $m_{1}^{y}m_{2}^{x}$, or
$m_{1}^{-x}m_{2}^{y}$.
\\
(4) Alice asks Bob if his measurements have been successful. If he
says no, then she goes to step 2. If he says yes, then she
considers the following two cases:

\noindent ($I$) Alice asks Bob if his measurement
belongs to set $A$. If he says no, then she goes to ($II$). If
he says yes (but of course he does not tell her along
which axes he performed his measurements),
and if she has chosen one of the four state
$\mid\!\alpha\rangle$, or $\mid\!\alpha^\prime\rangle$,
or $\mid\!\beta\rangle$, or $\mid\!\beta^\prime\rangle$,
then she tells him that
the protocol has not been successful,
and she goes to step $(2)$. But
if she has
chosen one of the four states
$\mid\!\phi\rangle$, or $\mid\!\phi^\prime\rangle$,
or $\mid\!\psi\rangle$, or
$\mid\!\psi^\prime\rangle$, then she tells him only
one of the following four alternatives:
\\
($i$) $\lambda$ is encoded in
$m_{1}^{x}m_{2}^{x}$,
\\
($ii$) $\lambda$ is encoded in
$-m_{1}^{x}m_{2}^{x}$,
\\
($iii$) $\lambda$ is encoded in
$m_{1}^{y}m_{2}^{y}$,
\\
($iv$) $\lambda$ is encoded in
$-m_{1}^{y}m_{2}^{y}$.

\noindent ($II$) Alice asks Bob if his measurement belongs to set $B$.
If he says yes, and
if she has chosen one of the four state
$\mid\!\phi\rangle$, or $\mid\!\phi^\prime\rangle$,
or $\mid\!\psi\rangle$, or
$\mid\!\psi^\prime\rangle$, then she tells him
the protocol has not been successful, and she goes to step $(2)$. But
if she has
chosen one of the four states
$\mid\!\alpha\rangle$, or $\mid\!\alpha^\prime\rangle$,
or $\mid\!\beta\rangle$, or $\mid\!\beta^\prime\rangle$,
then she tells him only
one of the following four alternatives:
\\
($i$) $\lambda$ is encoded in
$m_{1}^{y}m_{2}^{x}$,
\\
($ii$) $\lambda$ is encoded in
$-m_{1}^{y}m_{2}^{x}$,
\\
($iii$) $\lambda$ is encoded in
$m_{1}^{-x}m_{2}^{y}$,
\\
($iv$) $\lambda$ is encoded in
$-m_{1}^{-x}m_{2}^{y}$.

\begin{em}Theorem\end{em}: Assuming that Bob can not store the bit
until step $4$ (when Alice asks him whether his measurement belongs
to set $A$ or $B$), the above
oblivious transfer protocol is secure
even against cheater with unlimited computing power.

\begin{em}Proof\end{em}: First we consider Alice's strategy. We
assume that Alice has chosen one of the four state
$\mid\!\phi\rangle$, or $\mid\!\phi^\prime\rangle$,
or $\mid\!\psi\rangle$, or $\mid\!\psi^\prime\rangle$,
and Bob's measurement belongs
to set $A$, i.e., he has measured
$m_{1}^{x}m_{2}^{x}$ or $m_{1}^{y}m_{2}^{y}$.
Alice should consider the following four cases:
\\
$(1)$ First assume $\lambda=1 (-1)$, and
Alice decides to encode $\lambda$ in
$m_{1}^{x}m_{2}^{x} \left(-m_{1}^{x}m_{2}^{x}\right)$.
In this case, Alice should  choose either $\mid\!\phi\rangle$
or $\mid\!\psi\rangle$, since for both of these states
$m_{1}^{x}m_{2}^{x}=1$.
If Bob measures the spin of the two particles along the $x$-axis, then
he learns the value of $\lambda$.
However, if he measures the spins of the particles
along the $y$-axis, then he
does not learn any information about $\lambda$, since
$\langle\phi\mid\sigma_{1}^{y}\sigma_{2}^{y}
 \mid\phi\rangle=-1$,
but
$\langle\psi\mid\sigma_{1}^{y}\sigma_{2}^{y}
\mid\psi\rangle=1$.
\\
$(2)$  Next assume  $\lambda=-1 (1)$, and
Alice decides to encode $\lambda$ in
$m_{1}^{x}m_{2}^{x} \left(-m_{1}^{x}m_{2}^{x} \right)$.
In this case, Alice should  choose either $\mid\!\phi^\prime\rangle$
or $\mid\!\psi^\prime\rangle$, since for both of these states
$m_{1}^{x}m_{2}^{x}=-1$.
If Bob measures the spin of the two particles along the $x$-axis, then
he learns the value of $\lambda$.
However, if he measures the spins of the particles
along the $y$-axis, then he
does not learn any information about $\lambda$, since
$\langle\phi^\prime\mid\sigma_{1}^{y}\sigma_{2}^{y}
\mid\phi^\prime\rangle=1$,
but
$\langle\psi^\prime\mid\sigma_{1}^{y}\sigma_{2}^{y}
\mid\psi^\prime\rangle=-1$.
\\
$(3)$ Next assume $\lambda=1 (-1)$, and
Alice decides to encode $\lambda$ in
$m_{1}^{y}m_{2}^{y} \left(-m_{1}^{y}m_{2}^{y}\right)$.
In this case, Alice should choose either
$\mid\!\phi^\prime\rangle$ or $\mid\!\psi\rangle$,
since for both of these states
$m_{1}^{y}m_{2}^{y} =1$
If Bob measures the spin of both particles along the $y$-axis, then
he learns the value of $\lambda$.
However, if he measures the spins of the particles
along the $x$-axis, then does not learn any information about
the value of $\lambda$, since
$\langle\phi^\prime\mid\sigma_{1}^{x}\sigma_{2}^{x}
\mid\phi^\prime\rangle=-1$,
but
$\langle\psi\mid\sigma_{1}^{x}\sigma_{2}^{x}
\mid\psi\rangle=1$.
\\
$(4)$ Finally assume $\lambda=-1 (1)$, and
Alice decides to encode $\lambda$ in
$m_{1}^{y}m_{2}^{y}\left(-m_{1}^{y}m_{2}^{y}\right)$.
In this case, Alice should choose either
$\mid\!\phi\rangle$ or $\mid\!\psi^\prime\rangle$,
since for both of these states
$m_{1}^{y}m_{2}^{y} =-1$
If Bob measures the spin of both particles along the $y$-axis, then
he learns the value of $\lambda$.
However, if he measures the spins of the particles
along the $x$-axis, then does not learn any information about
the value of $\lambda$, since
$\langle\phi\mid\sigma_{1}^{x}\sigma_{2}^{x}
\mid\phi\rangle=1$,
but
$\langle\psi^\prime\mid\sigma_{1}^{x}\sigma_{2}^{x}
\mid\psi^\prime\rangle=-1$.

We now consider Bob's strategy. If Bob is
honest, then the oblivious transfer
protocol can succeed without any difficulty (see above).
Consider now a cheating Bob who measures the spin of
the first particle along
axis $\vec{a}$ and measures the  spin of the second particle
along axis $\vec{b}$,
i.e., $m_{1}^{a}m_{2}^{b}$. Assume (without loss of
generality) that Bob obtains
$m_{1}^{a}m_{2}^{b} = 1$.
Bob then asks the following question:
Given that
$m_{1}^{a}m_{2}^{b} = 1$,
what is the probability that
$m_{1}^{x}m_{2}^{x} = 1$, i.e., what is
$p\;\left(m_{1}^{x}m_{2}^{x} =1\mid m_{1}^{a}m_{2}^{b}=1\right)$?
To answer this question, he notes that only states
$\mid\!\phi\rangle$ and $\mid\!\psi\rangle$ can contribute to
$m_{1}^{x}m_{2}^{x} =1$. Thus

\begin{eqnarray}
p\,\left(m_{1}^{x}m_{2}^{x} =1\mid m_{1}^{a}m_{2}^{b}=1\right)=
p\,\left(\mid\!\phi\rangle\mid m_{1}^{a}m_{2}^{b}=1\right)+
p\,\left(\mid\!\psi\rangle\mid m_{1}^{a}m_{2}^{b}=1\right).
\end{eqnarray}
To calculate
$p\;\left(\mid\!\phi\rangle\mid m_{1}^{a}m_{2}^{b}=1\right)$,
note that

\begin{eqnarray}
p\;\left(\mid\!\phi\rangle\; , m_{1}^{a}m_{2}^{b}=1\right)
& = &p\;\left(\mid\!\phi\rangle\right)\:\:
p\;\left (m_{1}^{a}m_{2}^{b} = 1\mid \;\,\, \mid\!\phi\rangle\right),
\nonumber\\
& = & p\;\left(m_{1}^{a}m_{2}^{b} = 1\right)\;
p\; \left(\mid\!\phi\rangle  \mid m_{1}^{a}m_{2}^{b} = 1\right).
\end{eqnarray}
But
\begin{eqnarray}
p\; \left(\mid\!\phi\rangle\right)&=&\frac{1}{4},\\
p\;\left(m_{1}^{a}m_{2}^{b} =1\right)&=&\frac{1}{2},\\
p\;\left(m_{1}^{a}m_{2}^{b} =1 \mid  \;\,\, \mid\!\phi\rangle\right)
&=&\cos^{2}
\left(\frac{\theta_{a}+\theta_{b}}{2}\right),
\end{eqnarray}
where (5) follows from the fact that Alice chooses any state with
probability $\frac{\displaystyle 1}{\displaystyle 4}$,
(6) follows from the symmetry of the
problem, and (7) follows from the standard rules of quantum mechanics.
\\
Substituting the above formulas in Eq. (4), we obtain

\begin{eqnarray}
p\; \left(\mid\!\phi\rangle\mid m_{1}^{a}m_{2}^{b} = 1\right)
=\frac{1}{2}\cos^{2}
\left(\frac{\theta_{a}+\theta_{b}}{2}\right).
\end{eqnarray}
\\
Similar argument shows that

 \begin{eqnarray}
p\; \left(\mid\!\psi\rangle\mid m_{1}^{a}m_{2}^{b} = 1\right)
=\frac{1}{2}\cos^{2}\left(\frac{\theta_{a}-\theta_{b}}{2}\right).
\end{eqnarray}
\\
Thus

\begin{eqnarray}
p\;\left(m_{1}^{x}m_{2}^{x} =1\mid m_{1}^{a}m_{2}^{b}=1\right)=
\frac{1}{2}\cos^{2}\left(\frac{\theta_{a}+\theta_{b}}{2}\right)+
\frac{1}{2}\cos^{2}\left(\frac{\theta_{a}-\theta_{b}}{2}\right).
\end{eqnarray}

Bob now asks the following question: Given that
$m_{1}^{a}m_{2}^{b} = 1$,
what is the probability that
$m_{1}^{y}m_{2}^{y} = 1$, i.e., what is
$p\;\left(m_{1}^{y}m_{2}^{y} =1\mid m_{1}^{a}m_{2}^{b}=1\right)$?
To answer this question, he notes that only states
$\mid\!\phi^\prime\rangle$ and $\mid\!\psi\rangle$ can contribute to
$m_{1}^{y}m_{2}^{y} =1$. Thus

\begin{eqnarray}
p\left(m_{1}^{y}m_{2}^{y} =1\mid m_{1}^{a}m_{2}^{b}=1\right)=
p\;\left(\mid\phi^\prime\rangle\mid m_{1}^{a}m_{2}^{b}=1\right)+
p\;\left(\mid\psi\rangle\mid m_{1}^{a}m_{2}^{b}=1\right).
\end{eqnarray}
\\
Similar argument as before shows that

\begin{eqnarray}
p\;\left(m_{1}^{y}m_{2}^{y} =1\mid m_{1}^{a}m_{2}^{b}=1\right)
=\frac{1}{2}\sin^{2}\left(\frac{\theta_{a}+\theta_{b}}{2}\right)+
\frac{1}{2}\cos^{2}\left(\frac{\theta_{a}-\theta_{b}}{2}\right).
\end{eqnarray}
\\
Now the probability that Bob learns the value of $\lambda$ is
$\frac{\displaystyle 1}{\displaystyle 2}
\;
\bigl[\; p\;(m_{1}^{y}m_{2}^{y} =1\mid m_{1}^{a}m_{2}^{b}=1) +
p\;(m_{1}^{x}m_{2}^{x} =1\mid m_{1}^{a}m_{2}^{b}=1)\;\bigr]$.
{}From Eqs. (12) and (14), we have

\begin{eqnarray}
\frac{1}{2}
\biggl[p\;\left(m_{1}^{y}m_{2}^{y}
=1\mid m_{1}^{a}m_{2}^{b}=1\right) &+&
p\;\left(m_{1}^{x}m_{2}^{x} =
1\mid m_{1}^{a}m_{2}^{b}=1\right)\biggr]=
\nonumber\\
\frac{\displaystyle1}{\displaystyle4}
&+&\;\;\frac{1}{2} \cos^{2} \left
(\frac{\theta_{a}-\theta_{b}}{2}\right).
\end{eqnarray}
\\
Note that the maximum value of the RHS of (13) is
$\frac{\displaystyle 3}{\displaystyle 4}$.
Thus the best strategy for Bob is to
measure the spins of both particles either along
the same axis. In particular, if Bob does not cheat and
measures the spins of both particles either along the $x$-axis or
along the $y$-axis, then he obtains maximum information about
the value of the OT bit $\lambda$.

Having demonstrated that if Alice chooses one of the four states
$\mid\!\phi\rangle$, or $\mid\!\phi^\prime\rangle$,
or $\mid\!\psi\rangle$, or $\mid\!\psi^\prime\rangle$,
and if Bob's measurement belongs to set $A$, then the
OT protocol can be implemented successfully,
we now consider the other alternative. We assume
that Alice has chosen one of the four state
$\mid\!\alpha\rangle$, or $\mid\!\alpha^\prime\rangle$,
or $\mid\!\beta\rangle$, or $\mid\!\beta^\prime\rangle$,
and Bob's measurement belongs to
set $B$, i.e., he has measured
$m_{1}^{y}m_{2}^{x}$, or $m_{1}^{-x}m_{2}^{y}$.
Again Alice should
consider the following four cases:

\noindent $(1)$ First assume $\lambda=1 (-1)$, and
Alice encodes $\lambda$ in
$m_{1}^{y}m_{2}^{x} \left(-m_{1}^{y}m_{2}^{x}\right)$.
In this case, Alice should  choose either $\mid\!\alpha\rangle$
or $\mid\!\beta\rangle$, since for both of these states
$m_{1}^{y}m_{2}^{x}=1$.
If Bob measures the spin of the first
particle along the $y$ axis, and spin of the second
particle along the $x$ axis, then
he learns the value of $\lambda$.
However, if he measures the spins of the first particle
along the $-x$ axis, and the spin of the second
particle along $y$ axis, then he
does not learn any information about $\lambda$, since
$\langle\alpha\mid\sigma_{1}^{-x}\sigma_{2}^{y}
 \mid\alpha\rangle=-1$,
but
$\langle\beta\mid\sigma_{1}^{-x}\sigma_{2}^{y}
\mid\beta\rangle=1$.
\\
$(2)$  Next assume  $\lambda=-1 (1)$, and
Alice encodes $\lambda$ in
$m_{1}^{y}m_{2}^{x} \left(-m_{1}^{y}m_{2}^{x} \right)$.
In this case, Alice should
choose either $\mid\!\alpha^\prime\rangle$
or $\mid\!\beta^\prime\rangle$, since for both of these states
$m_{1}^{y}m_{2}^{x}=-1$.
If Bob measures the spin of the first
particle along the $y$ axis, and spin of the second
particle along the $x$ axis, then
he learns the value of $\lambda$.
However, if he measures the spins of the first particle
along the $-x$ axis, and the spin of the second
particle along $y$ axis, then he
does not learn any information about $\lambda$, since
$\langle\alpha^\prime\mid\sigma_{1}^{-x}\sigma_{2}^{y}
\mid\alpha^\prime\rangle=1$,
but
$\langle\beta^\prime\mid\sigma_{1}^{-x}\sigma_{2}^{y}
\mid\beta^\prime\rangle=-1$.
\\
$(3)$ Next assume $\lambda=1 (-1)$, and
Alice encodes $\lambda$ in
$m_{1}^{-x}m_{2}^{y} \left(-m_{1}^{-x}m_{2}^{y}\right)$.
In this case, Alice should choose either
$\mid\!\alpha^\prime\rangle$ or $\mid\!\beta\rangle$,
since for both of these states
$m_{1}^{-x}m_{2}^{y} =1$
If Bob measures the spin of the first particle along
-$x$ axis and spin of the second particle along the $y$-axis, then
he learns the value of $\lambda$.
However, if he measures the spin of  of the first particle
along the $y$ axis and spin of the second particle along
the $x$ axis, then
he does not learn any information about
the value of $\lambda$, since
$\langle\alpha^\prime\mid\sigma_{1}^{y}\sigma_{2}^{x}
\mid\alpha^\prime\rangle=-1$,
but
$\langle\beta\mid\sigma_{1}^{y}\sigma_{2}^{x}
\mid\beta\rangle=1$.
\\
$(4)$ Finally assume $\lambda=-1 (1)$, and
Alice encodes $\lambda$ in
$m_{1}^{-x}m_{2}^{y}\left(-m_{1}^{-x}m_{2}^{y}\right)$.
In this case, Alice should choose either
$\mid\!\alpha\rangle$ or $\mid\!\beta^\prime\rangle$,
since for both of these states
$m_{1}^{-x}m_{2}^{y} =-1$
If Bob measures the spin of both particles along the $y$-axis, then
he learns the value of $\lambda$.
However, if he measures the spin of  of the first particle
along the $y$ axis and spin of the second particle along
the $x$ axis, then
he does not learn any information about
the value of $\lambda$, since
$\langle\alpha\mid\sigma_{1}^{y}\sigma_{2}^{x}
\mid\alpha\rangle=-1$,
but
$\langle\beta^\prime\mid\sigma_{1}^{y}\sigma_{2}^{x}
\mid\beta^\prime\rangle=1$.

If Bob is honest, then the OT
protocol can succeed without any difficulty (see above).
However, if Bob is dishonest, the same argument as before
shows that he does not gain any additional information by cheating.
Thus the OT protocol is secure even against cheaters with unlimited
computing power.

There is another flavor of OT which is
called one-out-of-two oblivious transfer. The goal of this
protocol is:

\noindent 1 - Alice has two bits $\lambda_{1}$ and
$\lambda_{2}$ where $\lambda_{1}$ (or $\lambda_{2}$)
is either $1$ or $-1$.

\noindent 2 - Bob chooses to obtain either bit $\lambda_1$ or
$\lambda_2$.

\noindent 3 - Bob knows whether or not he has obtained the bit.

\noindent 4 - Alice does not learn which  bit Bob has chosen.

\noindent Less formally, Alice has two bits. Bob can get only one
of them, and Alice does not learn which bit Bob obtained.

This protocol can be implemented by

\noindent(1) Alice and Bob agree
that $\lambda_1$ is encoded in
\begin{math}
m_{1}^{x}m_{2}^{x},
\end{math}
or
\begin{math}
-m_{1}^{x}m_{2}^{x},
\end{math}
or
\begin{math}
m_{1}^{y}m_{2}^{y},
\end{math}
or
\begin{math}
-m_{1}^{y}m_{2}^{y},
\end{math}
and $\lambda_2$ is encoded in
\begin{math}
m_{1}^{y}m_{2}^{x},
\end{math}
or
\begin{math}
-m_{1}^{y}m_{2}^{x},
\end{math}
or
\begin{math}
m_{1}^{-x}m_{2}^{y},
\end{math}
or
\begin{math}
-m_{1}^{-x}m_{2}^{y}.
\end{math}
They also agree
on a set $A$ defined as
$A=\{m_{1}^{x}m_{2}^{x},m_{1}^{y}m_{2}^{y}\}$, and on a set $B$
defined as
$B=\{{m_{1}^{y}m_{2}^{x},m_{1}^{-x}m_{2}^{y}}\}$.
\\
(2) Alice encodes $\lambda_1$ in
\begin{math}
m_{1}^{x}m_{2}^{x},
\end{math}
or
\begin{math}
-m_{1}^{x}m_{2}^{x},
\end{math}
or
\begin{math}
m_{1}^{y}m_{2}^{x},
\end{math}
or
\begin{math}
-m_{1}^{y}m_{2}^{x},
\end{math}
and $\lambda_2$ in
\begin{math}
m_{1}^{y}m_{2}^{y},
\end{math}
or
\begin{math}
-m_{1}^{y}m_{2}^{y},
\end{math}
or
\begin{math}
m_{1}^{-x}m_{2}^{y},
\end{math}
or
\begin{math}
-m_{1}^{-x}m_{2}^{y},
\end{math}
chosen randomly by her.
She randomly chooses an appropriate state
(as shown below, there are two appropriate states
for any of her choices).
She then prepares a pair of particles in that state and sends
both particles to Bob.
\\
(3) Bob measures randomly either $m_{1}^{x}m_{2}^{x}$ or
$m_{1}^{y}m_{2}^{y}$, or $m_{1}^{y}m_{2}^{x}$, or
$m_{1}^{-x}m_{2}^{y}$.
\\
(4) Alice asks Bob if his measurements have been successful. If he
says no, then she goes to step 2. If he says yes, then she
considers the following two cases:

\noindent ($I$) Alice asks Bob if his measurement
belongs to set $A$. If he says no, then Alice goes to step ($II$).
If he says yes (but of course he does not tell her along which
axes he performed his measurements),
and if she has chosen one of the four state
$\mid\!\alpha\rangle$, or $\mid\!\alpha^\prime\rangle$,
or $\mid\!\beta\rangle$, or $\mid\!\beta^\prime\rangle$,
then she tells him that
the protocol has not been successful, and she goes to step $(2)$. But
if Alice has
chosen one of the four states
$\mid\!\phi\rangle$, or $\mid\!\phi^\prime\rangle$,
or $\mid\!\psi\rangle$, or $\mid\!\psi^\prime\rangle$,
then she tells him
that $\lambda_1$ is encoded in
\begin{math}
m_{1}^{x}m_{2}^{x}
\end{math}
or
\begin{math}
-m_{1}^{x}m_{2}^{x},
\end{math}
and $\lambda_2$ is encoded in
\begin{math}
m_{1}^{y}m_{2}^{y}
\end{math}
or
\begin{math}
-m_{1}^{y}m_{2}^{y}.
\end{math}

\noindent ($II$) Alice asks Bob if his measurement
belongs to set $B$. If
he says yes, and if she has chosen one of the four state
$\mid\!\phi\rangle$, or $\mid\!\phi^\prime\rangle$,
or $\mid\!\psi\rangle$, or $\mid\!\psi^\prime\rangle$,
then she tells him that
the protocol has not been successful, and she goes to step $(2)$. But
if Alice has
chosen one of the four states
$\mid\!\alpha\rangle$, or $\mid\!\alpha^\prime\rangle$,
or $\mid\!\beta\rangle$, or $\mid\!\beta^\prime\rangle$,
then she tells him
that $\lambda_1$ is encoded in
\begin{math}
m_{1}^{y}m_{2}^{x}
\end{math}
or
\begin{math}
-m_{1}^{y}m_{2}^{x},
\end{math}
and $\lambda_2$ is encoded in
\begin{math}
m_{1}^{-x}m_{2}^{y}
\end{math}
or
\begin{math}
-m_{1}^{-x}m_{2}^{y}.
\end{math}

\begin{em}Theorem\end{em}: Assuming that Bob can not store the bit
until step $4$ (when Alice asks him whether his measurement belongs
to set $A$ or $B$), the above
one-out-of-two oblivious transfer protocol is secure
against cheater with unlimited computing power.

\begin{em}Proof\end{em}:
First Assume that Bob performed his measurement in set $A$, and
Alice has chosen one of the four states
$\mid\!\phi\rangle$, or $\mid\!\phi^\prime\rangle$,
or $\mid\!\psi\rangle$, or $\mid\!\psi^\prime\rangle$,
(similar argument
also applies if Alice has chosen one of the four states
$\mid\!\alpha\rangle$, or $\mid\!\alpha^\prime\rangle$,
or $\mid\!\beta\rangle$, or $\mid\!\beta^\prime\rangle$).
She should consider the following four cases:

\noindent $(i)$ First assume  $\lambda_1=1$, $\lambda_2=1$,
and Alice decides to encode $\lambda_1$ in
$m_{1}^{x}m_{2}^{x}$, and $\lambda_2$ in
$m_{1}^{y}m_{2}^{y}$, or
$\lambda_1=-1$, $\lambda_2=-1$,
and Alice decides to encode $\lambda_1$ in
$-m_{1}^{x}m_{2}^{x}$, and $\lambda_2$ in
$-m_{1}^{y}m_{2}^{y}$
In this case, Alice should choose
state $\mid\!\psi\rangle$, since for this state
$m_{1}^{x}m_{2}^{x}=1$, and
$m_{1}^{y}m_{2}^{y}=1$.
\\
$(ii)$ Next assume  $\lambda_1=1$, $\lambda_2=-1$,
and Alice decides to encode $\lambda_1$ in
$m_{1}^{x}m_{2}^{x}$, and $\lambda_2$ in
$m_{1}^{y}m_{2}^{y}$, or
$\lambda_1=-1$, $\lambda_2=1$,
and Alice decides to encode $\lambda_1$ in
$-m_{1}^{x}m_{2}^{x}$, and $\lambda_2$ in
$-m_{1}^{y}m_{2}^{y}$.
In this case, Alice should choose
state $\mid\!\phi\rangle$, since for this state
$m_{1}^{x}m_{2}^{x}=1$, and
$m_{1}^{y}m_{2}^{y}=-1$.
\\
$(iii)$ Next assume  $\lambda_1=-1$, $\lambda_2=1$,
and Alice decides to encode $\lambda_1$ in
$m_{1}^{x}m_{2}^{x}$, and $\lambda_2$ in
$m_{1}^{y}m_{2}^{y}$, or
$\lambda_1=1$, $\lambda_2=-1$,
and Alice decides to encode $\lambda_1$ in
$-m_{1}^{x}m_{2}^{x}$, and $\lambda_2$ in
$-m_{1}^{y}m_{2}^{y}$.
In this case, Alice should choose
state $\mid\!\phi^\prime\rangle$, since for this state
$m_{1}^{x}m_{2}^{x}=-1$, and
$m_{1}^{y}m_{2}^{y}=1$.
\\
$(iv)$ Finally assume  $\lambda_1=-1$, $\lambda_2=-1$,
and Alice decides to encode $\lambda_1$ in
$m_{1}^{x}m_{2}^{x}$, and $\lambda_2$ in
$m_{1}^{y}m_{2}^{y}$, or
$\lambda_1=1$, $\lambda_2=1$,
and Alice decides to encode $\lambda_1$ in
$-m_{1}^{x}m_{2}^{x}$, and $\lambda_2$ in
$-m_{1}^{y}m_{2}^{y}$.
In this case, Alice should choose
state $\mid\!\psi^\prime\rangle$, since for this state
$m_{1}^{x}m_{2}^{x}=-1$, and
$m_{1}^{y}m_{2}^{y}=-1$.

The same argument that was used for quantum OT can be used
to prove that Bob does not
gain any additional information by setting his polarizer at other
angles.
Thus the one-out-of-two OT protocol
is secure even against cheaters
with unlimited computing power.
\pagebreak

\end{document}